\documentclass[12pt]{article}

\usepackage{amsmath,amssymb,bm}
\usepackage{mathbbol}
\usepackage{graphicx}
\usepackage{xcolor}
\definecolor{npblue}{HTML}{1F4E9C}
\definecolor{ppred}{HTML}{B02A2A}
\usepackage{booktabs}
\usepackage{float}

\usepackage{braket}
\usepackage{geometry}
\usepackage[
colorlinks=true,
linkcolor=blue,
citecolor=blue,
urlcolor=blue,
breaklinks=true
]{hyperref}

\title{\textbf{The Fock representation of the Coulomb interaction.\\ II. Three-body bound states}}
\author{M.~M.~Nishonov\thanks{Email: m.nishonov@nuu.uz}\\
	National University of Uzbekistan, 100174 Tashkent, Uzbekistan}
\date{}

\begin{document}
\maketitle

\begin{abstract}
The Fock representation of the unscreened Coulomb interaction, a block of separable terms with analytic form factors and strengths, is included in the momentum-space Faddeev kernel of three-body bound states beside the separable nuclear terms. A bound-state kernel evaluates every pair interaction at negative pair energies, where the representation is exact and convergent, so the Coulomb force enters without a screening radius and without a limiting procedure. With this representation the $^{3}$H--$^{3}$He splitting of the CD~Bonn potential is in very good agreement with the exact Faddeev result; in the $S$-wave model space the Coulomb-distorted form factors give about half the Fock value. The Feshbach--Schur projection of the Pauli-forbidden states acts on the forbidden eigenstates of the nuclear-plus-Coulomb pair Hamiltonian, the Coulomb-dressed states obtained inside the extended separable problem, as the eigenstate condition requires. The Coulomb displacements of the cluster-model states of $^{6}$Li, $^{9}$Be and $^{12}$C and of the $^{9}$Be--$^{9}$B mirror pair are given. The eigenstate condition is tested in $^{9}$Be and $^{12}$C by replacing the Coulomb-dressed forbidden states with those of the Coulomb-subtracted interaction and with the oscillator functions of the orthogonality-condition model.
\end{abstract}

\section{Introduction}
\label{sec:intro}

The three-nucleon system and the light nuclei which the cluster model describes as three bodies, $^{6}$He and $^{6}$Li as $\alpha+N+N$, $^{9}$Be and $^{9}$B as $\alpha+\alpha+N$ and $^{12}$C as $\alpha+\alpha+\alpha$, are the simplest systems in which the Coulomb force between charged pairs acts inside a three-body kernel. The charged pairs are the $pp$ pair in $^{3}$He, the $\alpha p$ pair in $^{6}$Li and $^{9}$Be, the $\alpha\alpha$ pair in $^{9}$Be and in $^{9}$B, where two $\alpha p$ pairs are added, and the three $\alpha\alpha$ pairs in $^{12}$C. In momentum-space Faddeev calculations with separable two-body input the Coulomb force has been treated by screening, by Coulomb-distorted form factors and by the Coulomb--Sturmian expansion. In the screening method the Coulomb potential is replaced by a screened one at a finite radius; in scattering the screening is removed afterwards by the renormalization of Ref.~\cite{Alt1978}, as in the proton--deuteron calculations of Refs.~\cite{Alt2002,Deltuva2005}, while a bound-state energy needs no renormalization and is followed to a radius at which it no longer changes. In the distortion method the nuclear form factors are dressed with the M{\o}ller wave operator of the screened Coulomb force, and the Coulomb-distorted form factors so obtained keep the separable structure. Analytic partial-wave Coulomb $t$~matrices at negative energies, obtained by Fock's method for integer values of the Coulomb parameter, are given in Refs.~\cite{Kharchenko2018,Kharchenko2017a,Kharchenko2017b,Kharchenko2019}. The exact Coulomb $t$~matrix was used in a momentum-space three-body bound-state calculation by Kok and van Haeringen~\cite{vanHaeringen1982}, in a trinucleon model.

A bound-state kernel samples only negative pair energies, and the three existing treatments are suited to that domain in different degrees. Screening was devised for scattering, where the singularity of the Coulomb matrix element at zero momentum transfer must be treated on shell; carried into a bound-state kernel it adds a non-separable, radius-dependent tail which is superfluous there. In the distortion method the Coulomb force enters only through form factors which are approximate and still carry the screening radius of the M{\o}ller operator. In the Coulomb--Sturmian expansion of Papp and Plessas~\cite{PappPlessas1996} the Coulomb potential and the Green's operators, not the pair interaction itself, are expanded in the Sturmian functions, which are the Fock modes of the present representation up to normalization. With none of the three is the Coulomb force of every charged pair exact and separable in the kernel. In Ref.~\cite{FockI} the stereographic projection of Fock~\cite{Fock1935} was used to write the unscreened Coulomb interaction of a charged pair as an exactly separable potential at negative energies, with analytic form factors and strengths, which enters the separable nuclear input as a further block of terms.  No screening radius, renormalization or distorted form factor appears; only the rank of the pair interaction increases, while the kernel and the recoupling of the purely nuclear problem are unchanged. Two-body benchmarks against the screening and distortion methods were given there, together with the relation of the representation to the analytic $t$~matrices and to the Coulomb--Sturmian expansion.

In cluster models the Pauli principle forbids certain states of the relative motion of two clusters, the $0s$ state of the $\alpha N$ pair and the $0s$, $1s$ and $0d$ states of the $\alpha\alpha$ pair. The orthogonality-condition model~\cite{Saito1969} removes them from the pair interaction, in practice by the orthogonalizing pseudopotential of Ref.~\cite{Kukulin1974}, a projector onto the forbidden state with a large strength $\lambda$. Exact eliminations of $\lambda$ were made for one forbidden state per pair in the $3\alpha$ Faddeev equations~\cite{Fujiwara2004OCM} and in configuration space~\cite{Schellingerhout1993}.  In Ref.~\cite{Nishonov2026a} the elimination was written for a separable representation, in which the forbidden state is one more form factor. The limit $\lambda\to\infty$ of the pair $t$~matrix is then a Schur complement over the forbidden block, valid for any rank and number of forbidden states and pairs, in the same basis as the interaction; this is the Feshbach--Schur projection of Ref.~\cite{Nishonov2026a}, and it removes exactly the forbidden bound state only when the projected vector is an eigenstate of the pair Hamiltonian in use, the eigenstate condition. The point of the present paper is the combination of this projection with the Fock block. Since the Coulomb force is part of the same separable problem, the pair Hamiltonian which the kernel represents contains it, its forbidden eigenstates are Coulomb-dressed, and the eigenstate condition can be met for charged pairs, which a treatment with a screened Coulomb force meets only for the screened pair Hamiltonian and a distorted one not at all. This was shown in Ref.~\cite{FockI} for the two-body problem, where the condition requires the projected vector to be the Coulomb-dressed forbidden state, the eigenstate of the extended separable problem in the basis of the nuclear and the Fock form factors.

Here this representation is placed in the bound-state kernel, with three results. First, the Coulomb displacements of $^{3}$He, $^{6}$Li, $^{9}$Be, $^{9}$B and $^{12}$C are computed from one separable input, the CD~Bonn nucleon--nucleon~\cite{Machleidt2001}, the KKNN $\alpha N$~\cite{KKNN1979} and the BFW $\alpha\alpha$~\cite{BFW1977} interactions in the separable representation of Ref.~\cite{NishonovValidation}. The $^{3}$H--$^{3}$He splitting of the CD~Bonn potential agrees with the exact Faddeev result to $3$~keV, while in the $S$-wave model space the Coulomb-distorted form factors give $0.37$ against $0.69$~MeV. Second, the Pauli-forbidden states of the charged pairs are projected as eigenstates of the pair Hamiltonian which that kernel contains, the Coulomb-dressed states which the Fock block provides inside the separable problem, several per pair and in several pairs at once. Third, the effect of the eigenstate condition on a three-body energy is obtained by replacing the dressed eigenstates with the oscillator forbidden states of the orthogonality-condition model~\cite{Saito1969} at the oscillator width of the relative motion.  With the Coulomb force on, the replacement changes $^{9}$Be by $0.42$~MeV and $^{12}$C by $2.29$~MeV. The three-body kernel with a separable, unscreened and convergent Coulomb force for every charged pair, the simultaneous projection of several Coulomb-dressed forbidden states per pair and the effect of the eigenstate condition on the binding energy of $^{9}$Be and on the Coulomb-dressed part of the $^{12}$C shift are the new elements of this work.

The extended separable problem in the kernel and the number of Fock modes it needs are described in Sec.~\ref{sec:fock}. In Sec.~\ref{sec:exact} the eigenstate condition is stated and the size of its effect in the three-cluster nuclei, where the projected pairs are charged pairs, is given. The three-body benchmarks follow in Sec.~\ref{sec:benchmarks}: the $^{3}$H--$^{3}$He splitting against the distortion method and the exact result, and the Coulomb displacements of $^{6}$Li, $^{9}$Be, the $^{9}$Be--$^{9}$B mirror pair and $^{12}$C. In Sec.~\ref{sec:prescription} the eigenstate condition is tested with the oscillator forbidden states in $A=6$, $^{9}$Be and $^{12}$C.  Section~\ref{sec:summary} summarizes the results and outlines the applications that follow. The convergence in the spectator mesh and in the Fock rank is documented in Appendix~\ref{sec:mesh}, and the state-by-state analysis of the oscillator comparison in Appendix~\ref{sec:appB}.

\section{The Fock block in the Faddeev kernel}
\label{sec:fock}

At negative pair energies the Coulomb force of a charged pair is represented by the Fock expansion of Ref.~\cite{FockI}: a sum of separable terms with the form factors $\Phi_{nL}(p)$, Gegenbauer polynomials in the stereographic variable of the pair momentum, and analytic strengths $\lambda_{nL}$, at a scale $\beta_{C}$ which controls the rate of convergence. The nuclear part of every pair is a separable potential of the Ernst--Shakin--Thaler type~\cite{EST1973} with the analytic form factors of Ref.~\cite{NishonovValidation}. The propagator integrals, the pair $\tau$~matrices and the exchange elements of the Faddeev kernel are built from the same functions.

\subsection{The extended basis and the rank}
\label{sec:kernel}

With separable input every pair channel is a set of form factors and a coupling matrix, and the Fock modes are added to it as further terms. The extended basis of a charged channel is
\begin{equation}
\bm{g} \;=\; \{\underbrace{\chi_{1},\ldots,\chi_{N_{r}}}_{\text{nuclear}},\;
\underbrace{\Phi_{0},\ldots,\Phi_{N_{C}-1}}_{\text{Fock}},\;
\underbrace{\phi_{1},\ldots,\phi_{N_{f}}}_{\text{forbidden}}\}.
\label{eq:basisorder}
\end{equation}
Here $\chi_{i}$, $i=1,\dots,N_{r}$, are the nuclear form factors of the channel (the notation of Ref.~\cite{Nishonov2026a}) and $N_{r}$ is its rank; $\Phi_{n}$, $n=0,\dots,N_{C}-1$, are the Fock modes of Ref.~\cite{FockI} in the wave of the channel, $N_{C}$ being the number of Fock modes per charged pair, called the Fock rank below; and $\phi_{f}$, $f=1,\dots,N_{f}$, are the forbidden states, each written as a Legendre sum over two poles (the two-pole form of Ref.~\cite{NishonovValidation}). The coupling matrix is block diagonal,
\begin{equation}
\tilde{\bm\Lambda}=\mathrm{diag}\bigl(\bm\Lambda,\{\lambda_{nL}\},\lambda\,\mathbf{1}_{N_{f}}\bigr),
\label{eq:Lamext}
\end{equation}
with the fitted nuclear block, the diagonal Fock block (the Coulomb block of Ref.~\cite{FockI}) with the analytic strengths of that work, and the forbidden block, whose entry $\lambda$ is the strength of the orthogonalizing pseudopotential. The strict projection is the limit $\lambda\to\infty$ of the pair $\tau$~matrix, taken in closed form (Sec.~\ref{sec:faddeev}). Every projection of this paper is strict unless a finite $\lambda$ is stated.

The rest of the kernel is unchanged. The propagator integrals are computed for the $\Phi_{n}$ as for the nuclear form factors, and the pair $\tau$~matrix is the inverse of $[\tilde{\bm\Lambda}^{-1}-\tilde{\bm{\mathcal D}}(z)]$ over the extended basis, $\tilde{\mathcal D}_{\rho\rho'}(z)=\braket{g_{\rho}|G_{0}(z)|g_{\rho'}}$ being the propagator matrix over that basis. In a coupled charged channel the Fock modes carry the lower orbital momentum of the pair, and the higher component receives no separate Fock modes (for the $pp$ $^{3}P_{2}$--$^{3}F_{2}$ channel, $L=1$ only).

Every tabulated energy of this paper, outside the rank tests of Table~\ref{tab:rank} and the $^{1}D_{2}$ check of Sec.~\ref{sec:a3}, uses $N_{C}=20$ Fock modes per charged pair, with $\beta_{C}=0.8$~fm$^{-1}$ for the $pp$ and $\alpha p$ pairs and $1.0$~fm$^{-1}$ for the $\alpha\alpha$ pairs (the two-body calculations of Ref.~\cite{FockI} used $1.0$~fm$^{-1}$ for $pp$). The scale affects only the rate of convergence in $N_{C}$, which is checked in the three-body calculations themselves (Appendix~\ref{sec:mesh}, Table~\ref{tab:rank}). The Fock rank is set by the two-body convergence sequences of Ref.~\cite{FockI}, in which the $\alpha p$ $0s$ state is converged at $N_{C}=20$ and the $\alpha\alpha$ dressed forbidden states are converged already at $N_{C}=20$.

\subsection{The Faddeev equations with the extended input}
\label{sec:faddeev}

The partition $a$ denotes the pair $(bc)$ with $a$ as the spectator; $\bm p_a$ and $\bm q_a$ are the corresponding Jacobi momenta, $\mu_a$ and $\bar{\mu}_a$ are the pair and the spectator reduced masses, and $E$ is the three-body energy. The partial-wave recoupling is given in full in Ref.~\cite{Nishonov2026a}. For pair interactions the wave function decomposes as
\begin{equation}
\Psi=\sum_a\Psi_a,\qquad
\Psi_a=G_0(E)\,t_a(E)\sum_{b\neq a}\Psi_b.
\label{eq:psidecomp}
\end{equation}
With separable pair input each component has the form
\begin{equation}
\Psi_a=G_0(E)\sum_{i=1}^{N_{r,a}}\ket{g^a_i}\,X^a_i(\bm q_a),
\label{eq:psicomp}
\end{equation}
with $N_{r,a}$ the number of form factors of the pair channel of partition $a$ (the extended rank for a charged pair), and the spectator amplitudes satisfy the Alt--Grassberger--Sandhas equations~\cite{AGS1967} in the quasiparticle form of Ref.~\cite{Lovelace1964}, homogeneous for a bound state,
\begin{equation}
X^a_i(\bm q_a)=\sum_{b\neq a}W_{ab}\,\zeta_{ab}\sum_{jk}
\int d^3q_b\;
\tau^a_{ij}\Bigl(E-\tfrac{q_a^2}{2\bar{\mu}_a}\Bigr)\,
\mathcal{Z}^{ab}_{jk}(\bm q_a,\bm q_b;E)\,X^b_k(\bm q_b),
\label{eq:ags}
\end{equation}
with the interaction-free exchange kernel
\begin{equation}
\mathcal{Z}^{ab}_{jk}(\bm q_a,\bm q_b;E)=
\frac{g^a_j\bigl(\bm p_a(\bm q_a,\bm q_b)\bigr)\,
      g^b_k\bigl(\bm p_b(\bm q_a,\bm q_b)\bigr)}
     {E-\dfrac{q_a^2}{2\mu_b}-\dfrac{q_b^2}{2\mu_a}
       -\dfrac{\bm q_a\!\cdot\!\bm q_b}{m_{c}}},
\label{eq:Zapp}
\end{equation}
where $m_c$ is the mass of the exchanged particle, $\bm\tau^a$ is the pair $t$~matrix $t_a$ of Eq.~(\ref{eq:psidecomp}) in the basis of its form factors, $W_{ab}$ are the multiplicity weights of the identical-particle reduction, and $\zeta_{ab}$ are the transposition phases~\cite{Nishonov2026a}. Both the Coulomb force and the Pauli projection enter Eq.~(\ref{eq:ags}) only through $\tau^a$.

For a charged pair the basis is the extended set~(\ref{eq:basisorder}), with the block-diagonal coupling matrix $\tilde{\bm\Lambda}$ of Eq.~(\ref{eq:Lamext}). The pair $\tau$~matrix is
\begin{equation}
\tilde{\bm\tau}(z)
=\bigl[\tilde{\bm\Lambda}^{-1}-\tilde{\bm{\mathcal D}}(z)\bigr]^{-1},
\qquad
\tilde{\mathcal D}_{\rho\rho'}(z)
=\braket{g_\rho|G_0(z)|g_{\rho'}},
\label{eq:tauext}
\end{equation}
where $\rho,\rho'$ run over the full extended basis~(\ref{eq:basisorder}), $N_{r}+N_C+N_f$ members, and $z=E-q_a^2/2\bar{\mu}_a$ is the pair energy at which Eq.~(\ref{eq:ags}) evaluates $\tau^a$.

The strict projection is the limit $\lambda\to\infty$ of Eq.~(\ref{eq:tauext}), taken as the Schur complement over the forbidden block~\cite{Nishonov2026a}. Writing $\mathsf{A}$ for the \emph{allowed} block, which here contains the nuclear and the Fock blocks together, and $\mathsf{F}$ for the forbidden one,
\begin{equation}
\bm\tau^{(\mathrm{eff})}(z)
=\Bigl[
\bm\Lambda_{\mathsf{A}}^{-1}
-\bm{\mathcal D}_{\mathsf{AA}}(z)
+\bm{\mathcal D}_{\mathsf{AF}}(z)\,
\bm{\mathcal D}_{\mathsf{FF}}^{-1}(z)\,
\bm{\mathcal D}_{\mathsf{FA}}(z)
\Bigr]^{-1},
\label{eq:taueff}
\end{equation}
with $\bm\Lambda_{\mathsf{A}}=\mathrm{diag}(\bm\Lambda,\{\lambda_{nL}\})$, $n=0,\dots,N_C-1$.  Equation~(\ref{eq:taueff}) is the allowed block of the limit; the limit of the full matrix~(\ref{eq:tauext}) has three further blocks,
\begin{equation*}
\tilde{\bm\tau}_{\mathsf{AF}}=-\bm\tau^{(\mathrm{eff})}\bm{\mathcal D}_{\mathsf{AF}}\bm{\mathcal D}_{\mathsf{FF}}^{-1},\qquad
\tilde{\bm\tau}_{\mathsf{FA}}=-\bm{\mathcal D}_{\mathsf{FF}}^{-1}\bm{\mathcal D}_{\mathsf{FA}}\bm\tau^{(\mathrm{eff})},\qquad
\tilde{\bm\tau}_{\mathsf{FF}}=-\bm{\mathcal D}_{\mathsf{FF}}^{-1}+\bm{\mathcal D}_{\mathsf{FF}}^{-1}\bm{\mathcal D}_{\mathsf{FA}}\bm\tau^{(\mathrm{eff})}\bm{\mathcal D}_{\mathsf{AF}}\bm{\mathcal D}_{\mathsf{FF}}^{-1},
\end{equation*}
through which the forbidden quasiparticle channels remain in the kernel with their exchange couplings; the strength of the pseudopotential has been eliminated analytically, so no large parameter appears. Equation~(\ref{eq:taueff}) is the exact $\lambda\to\infty$ limit for any $\phi_f$; the eigenstate condition of Sec.~\ref{sec:exact} fixes the $\phi_f$ as the eigenstates of the pair Hamiltonian which the $\mathsf{A}$~block represents, with the Fock block included, the Coulomb-dressed forbidden states (Table~\ref{tab:pairspec} of Appendix~\ref{sec:appB} shows the removal for the Coulomb-off $\alpha\alpha$ pair). The convergence of the finite-strength pseudopotential to this limit is shown in Table~\ref{tab:lambda}.

The forbidden block may contain several states of one channel and several channels of one pair, and $\bm{\mathcal D}_{\mathsf{FF}}$ is then a matrix inverted at every pair energy. For the $\alpha\alpha$ $L=0$ pair it is the $2\times2$ block of the $0s$ and $1s$ states, coupled by the propagator integral, with the $0d$ state as a separate $L=2$ block; in $^{9}$Be the $\alpha\alpha$ states are projected simultaneously with the $\alpha N$ $0s$ state of the other two partitions, the simultaneous projection formulated in Ref.~\cite{Nishonov2026a}. In the $3\alpha$ system the three $\alpha\alpha$ pairs are identical, so after the identical-particle reduction the projection acts in the single $\alpha\alpha$ partition and the multiplicity weights $W_{ab}$ of Eq.~(\ref{eq:ags}) distribute it over the three pairs. The complement is well conditioned: for the $\alpha\alpha$ $L=0$ pair the condition number of $\bm{\mathcal D}_{\mathsf{FF}}$ lies between $1.4$ and $9.3$ at pair energies from $-5$ to $-150$~MeV for every set of forbidden states used in this paper.

The four blocks of the limit replace $\tau^a$ in Eq.~(\ref{eq:ags}) for every pair with forbidden states. For a bound state every pair energy $z=E-q_a^{2}/2\bar{\mu}_a$ at which Eq.~(\ref{eq:ags}) samples $\bm\tau^{(\mathrm{eff})}$ is negative, the domain on which the Fock part of the allowed block $\mathsf{A}$ is a convergent separable representation of the unscreened Coulomb interaction (Ref.~\cite{FockI}); the whole Coulomb and projection content of the calculation is carried by the finite matrix inverted in Eq.~(\ref{eq:taueff}).

\subsection{Meshes and precision of the tabulated energies}
\label{sec:meshes}

The spectator momentum is discretized on a mapped Gauss--Legendre mesh of $N_{q}$ points and, for $^{12}$C, whose amplitude lies where the mapped mesh is sparse, also on a piecewise grid with $n_{g}$ points per interval; both are defined in Appendix~\ref{sec:mesh}. All differences quoted in the tables are formed from the unrounded energies. Table~\ref{tab:precision} gives the mesh at which each system is tabulated and the precision of its entries; the convergence in the mesh and in the Fock rank is documented in Appendix~\ref{sec:mesh}. Differences between two calculations on the same mesh converge before the energies do, and such differences are the quantities on which the eigenstate condition is tested in Secs.~\ref{sec:exact} and~\ref{sec:prescription}.

\begin{table}[htbp]
\caption{Mesh of the tabulated energies and their precision, from the convergence sequences of Appendix~\ref{sec:mesh}.}
\label{tab:precision}
\begin{center}
\begin{tabular}{lp{4.6cm}p{5.6cm}}
\toprule
system & mesh & precision of the entries \\
\midrule
$A=3$    & $N_{q}=32$ & splitting within a keV of its value at $N_{q}=16$ \\
$A=6$ & $N_{q}=22$, $26$ (Table~\ref{tab:lambda}); $32$ (Table~\ref{tab:3b}) & energies converged to a few keV at $N_{q}=26$ \\
$^{9}$Be & $N_{q}=32$, $N_{C}=20$ & within about $3$~keV of the $N_{q}=48$, $N_{C}=30$ value \\
$^{9}$B  & $N_{q}=32$ & changes by $3$~keV to $N_{q}=40$ \\
$^{12}$C & grid, $n_{g}=10$ & about $55$~keV below the converged grid value; rank residual $18$~keV at $n_{g}=10$, $3$~keV at $n_{g}=20$ \\
\bottomrule
\end{tabular}
\end{center}
\end{table}

\section{The eigenstate condition in the three-body problem}
\label{sec:exact}

The Feshbach--Schur elimination of a Pauli-forbidden state~\cite{Nishonov2026a}, the $\lambda\to\infty$ limit of the orthogonalizing pseudopotential~\cite{Kukulin1974} taken as a Schur complement, leaves the projected vector free, and the vector determines which state is eliminated~\cite{FockI,Nishonov2026a}. The eigenstate condition~\cite{Nishonov2026a} restricts that choice: only an eigenstate of the pair Hamiltonian present in the calculation removes exactly the forbidden state and leaves the scattering states of the pair unchanged. With the Fock block in the kernel that Hamiltonian contains the Coulomb force, so the projected states are the Coulomb-dressed forbidden states, the bound states of the nuclear-plus-Coulomb pair interaction, obtained directly as eigenstates of the extended separable problem (Ref.~\cite{FockI}, Sec.~3). They are the states projected everywhere in this paper outside the comparisons which test the condition.

The effect is largest in $^{9}$Be and $^{12}$C, where the forbidden pairs are charged pairs. With the $\alpha\alpha$ Coulomb force on, the projected states may be taken as the eigenstates of the dressed interaction, for which the condition holds, or of the separable potential with the Coulomb force subtracted, called the undressed states, for which it does not. The undressed states overlap the dressed ones at $0.998$ ($0s$), $0.995$ ($1s$) and $0.999$ ($0d$), the first-order-small admixture expected for a Coulomb shift of a few MeV on states bound by tens of MeV; the projected $\alpha\alpha$ pair keeps no deep state with either set, only the shallow $^{8}$Be state, a resonance with the Coulomb force on. Table~\ref{tab:dressedtest} compares the two choices with the same mesh, $\alpha\alpha$ separable potential and Fock block. In $^{9}$Be, with one $\alpha\alpha$ pair, the undressed calculation is the more bound by $60$~keV at every mesh from $N_{q}=22$ on, while the energy itself is still converging. In $^{12}$C, with three pairs, the two calculations differ by $1.6$--$1.7$~MeV at every mesh from $N_{q}=26$ on and by $1.63$~MeV on the $n_{g}=10$ grid of the tables, although the energy itself is not yet converged on the mapped mesh (Appendix~\ref{sec:mesh}).

\begin{table}[htbp]
\caption{$^{9}$Be and $^{12}$C ground-state energies (MeV) with the Fock block on the $\alpha\alpha$ pairs and the dressed or the undressed (Coulomb-subtracted) eigenstates projected, against the number $N_{q}$ of spectator-momentum mesh points; the last column is undressed minus dressed. The rows ``grid'' are the piecewise grid of Appendix~\ref{sec:mesh} with $n_{g}=10$ and $15$.}
\label{tab:dressedtest}
\begin{center}
\begin{tabular}{lcccc}
\toprule
system & $N_{q}$ & dressed & undressed & difference \\
\midrule
$^{9}$Be  & 22 & $-1.986$  & $-2.050$  & $-0.06$ \\
          & 26 & $-2.027$  & $-2.089$  & $-0.06$ \\
          & 30 & $-2.039$  & $-2.102$  & $-0.06$ \\
          & 32 & $-2.041$  & $-2.103$  & $-0.06$ \\
$^{12}$C  & 22 & $-23.13$  & $-21.24$  & $+1.90$ \\
          & 26 & $-21.18$  & $-19.51$  & $+1.66$ \\
          & 30 & $-20.54$  & $-18.89$  & $+1.65$ \\
          & 32 & $-20.42$  & $-18.77$  & $+1.65$ \\
          & 40 & $-20.05$  & $-18.42$  & $+1.63$ \\
          & grid, $n_{g}=10$ & $-19.96$ & $-18.34$  & $+1.63$ \\
          & grid, $n_{g}=15$ & $-19.92$ & $-18.30$  & $+1.62$ \\
\bottomrule
\end{tabular}
\end{center}
\end{table}

The $^{12}$C energy is a model quantity: the model with inert $\alpha$ particles overbinds the $3\alpha$ ground state by more than $10$~MeV, a documented property of such models~\cite{Fujiwara2004OCM,Matsumura2006}. Table~\ref{tab:fuj} compares the grid value with the calculation of Fujiwara, Kohno and Suzuki~\cite{Fujiwara2004OCM} with the same potential; the two numbers are not directly comparable, and their agreement is not a test. The $\alpha$ mass used here is four nucleon masses, $4m_{N}=3756.3$~MeV with $m_{N}=939.08$~MeV, the mass used in Ref.~\cite{FockI} and in the separable tables of Ref.~\cite{NishonovValidation}, against the physical mass of Ref.~\cite{Fujiwara2004OCM}; at the physical mass, with the separable potential and its Coulomb-dressed forbidden states constructed for that mass, the binding is $0.79$~MeV less (Table~\ref{tab:fuj}).  The remainder, of similar size and in the opposite direction, comes from two differences in the treatment.  Reference~\cite{Fujiwara2004OCM} cuts the Coulomb force off at $10$~fm inside the kernel and projects the bound states of that cut-off potential, whereas here the point Coulomb interaction enters through the Fock block, the difference between the folded and the point force being absorbed in the short-range part of the separable potential (Ref.~\cite{FockI}). The projected states are here the Coulomb-dressed eigenstates of the separable representation, whose deepest state lies a quarter of an MeV above the state of the original potential at the mass used here (Ref.~\cite{FockI}) and a tenth of an MeV above the state of the cut-off potential at the physical mass. For a state bound by $20$~MeV the cut-off of the Coulomb force should matter little, so the remainder comes mainly from the separable representation and from the projected states, and the agreement of the two calculations appears to be a near-cancellation of these two offsets.

\begin{table}[htbp]
\caption{The $3\alpha$ ground state with the Coulomb-dressed forbidden states on the piecewise grid, $n_{g}=10$, against the calculation of Ref.~\cite{Fujiwara2004OCM} with the same potential (MeV). The two calculations differ in the $\alpha$ mass and in the treatment of the Coulomb force and of the forbidden states, as described in the text.}
\label{tab:fuj}
\begin{center}
\begin{tabular}{lr}
\toprule
 & $E(^{12}\mathrm{C})$ \\
\midrule
this work, $m_{\alpha}=4m_{N}$                     & $-19.964$ \\
this work, physical $m_{\alpha}$                   & $-19.172$ \\
\quad physical minus $4m_{N}$                       & $+0.792$ \\
Ref.~\cite{Fujiwara2004OCM}, physical $m_{\alpha}$ & $-19.897$ \\
\quad Ref.~\cite{Fujiwara2004OCM} minus this work at the physical mass & $-0.725$ \\
experiment                                         & $-7.275$ \\
\bottomrule
\end{tabular}
\end{center}
\end{table}

\section{Three-body benchmarks}
\label{sec:benchmarks}

The two-body benchmarks of Ref.~\cite{FockI} compared the Fock block with the screening and the distortion method where the exact two-body solution exists.  The three-body benchmarks below are on quantities not fixed by two-body data: the Coulomb part of the $^{3}$H--$^{3}$He splitting and the Coulomb displacements of the three-cluster nuclei. The constants are those of Ref.~\cite{FockI} and the masses those of Sec.~\ref{sec:exact}. In the particle basis the nucleons are distinguished by charge, and the $\alpha p$, $\alpha n$ and the $pp$, $np$, $nn$ pairs carry separate interactions; in the isospin basis the $\alpha n$ and $np$ interactions represent all charge states. Outside the comparison of Table~\ref{tab:he3} and the $^{1}D_{2}$ test of Sec.~\ref{sec:a3}, every bound-state calculation of this paper treats the Coulomb force by the Fock block alone: the $j\le4$ $^{3}$He calculation carries it on the $pp$ $S$ and $P$ waves, all other calculations on every charged pair wave.

\subsection{The $^{3}$H--$^{3}$He splitting}
\label{sec:a3}

The quantity is the splitting $\Delta E=E(^{3}\mathrm{H})-E(^{3}\mathrm{He})$, computed with the CD~Bonn separable potentials and with the $pp$ Coulomb force treated by each method in turn; $E(^{3}\mathrm{H})$ is identical in both, because $^{3}$H has no charged pair. The Coulomb part of this splitting is a well-studied quantity~\cite{Brandenburg1988}, used here only as a check of the Coulomb treatment; about $70$~keV of the experimental value is charge-symmetry breaking~\cite{Brandenburg1988}, already contained in the charge-dependent separable potentials. Table~\ref{tab:he3} compares the two methods in the $S$-wave-pair model space, the $^{1}S_{0}$ and the coupled $^{3}S_{1}$--$^{3}D_{1}$ pair waves, six three-body channels for $J^{\pi}=1/2^{+}$ in the particle basis, at $N_{q}=16$.  With the Fock representation the splitting is $0.69$~MeV, with the distortion method $0.37$~MeV; the latter is insensitive to the screening radius and to whether the M{\o}ller operator is taken in first order or in full, and does not approach the Fock value. This splitting alone does not decide between the methods; the two-body comparison is made in Ref.~\cite{FockI}.

\begin{table}[htbp]
\caption{$^{3}$He binding energy and the splitting $\Delta E=E(^{3}\mathrm{H})-E(^{3}\mathrm{He})$ with the Fock block and with the distortion method, $S$-wave-pair model space at $N_{q}=16$ (MeV), with $E(^{3}\mathrm{H})=-8.0723$~MeV; in this space the $pp$ Coulomb force acts in the $^{1}S_{0}$ wave only; $R$ is the screening radius of the distortion method.}
\label{tab:he3}
\begin{center}
\begin{tabular}{lcc}
\toprule
calculation & $E(^{3}\mathrm{He})$ (MeV) & $\Delta E$ (MeV) \\
\midrule
Fock block on $pp$                   & $-7.3797$ & $0.693$ \\
distortion, $R=10$~fm                & $-7.7022$ & $0.370$ \\
distortion, $R=20$~fm                & $-7.7009$ & $0.371$ \\
distortion, $R=40$~fm                & $-7.7014$ & $0.371$ \\
distortion, $R=80$~fm                & $-7.7035$ & $0.369$ \\
distortion, first order ($R=20$)     & $-7.6931$ & $0.379$ \\
\bottomrule
\end{tabular}
\end{center}
\end{table}

\begin{table}[!htbp]
\caption{The $^{3}$H--$^{3}$He splitting in the Fock representation (MeV). The model space is the set of pair waves, the second column lists the $pp$ waves which carry the Fock block, and $N_{q}$ is the number of spectator-momentum mesh points. The exact Faddeev energies of the CD~Bonn potential are those of Ref.~\cite{Nogga2003}; the experimental energies~\cite{Wang2021} are relative to the three-nucleon threshold.}
\label{tab:a3}
\begin{center}
\begin{tabular}{llcccc}
\toprule
model space & $pp$ waves with Fock block & $N_{q}$ & $E(^{3}\mathrm{H})$ & $E(^{3}\mathrm{He})$ & $\Delta E$ \\
\midrule
$S$-wave pairs & $^{1}S_{0}$ & $16$ & $-8.0723$ & $-7.3797$ & $0.693$ \\
$j\le4$ & $^{1}S_{0}$ & $16$ & $-7.961$ & $-7.267$ & $0.694$ \\
$j\le4$ & $^{1}S_{0}$, $^{3}P_{0}$, $^{3}P_{1}$, $^{3}P_{2}$--$^{3}F_{2}$ & $16$ & $-7.961$ & $-7.226$ & $0.735$ \\
$j\le4$ & $^{1}S_{0}$ & $32$ & $-7.964$ & $-7.270$ & $0.694$ \\
$j\le4$ & $^{1}S_{0}$, $^{3}P_{0}$, $^{3}P_{1}$, $^{3}P_{2}$--$^{3}F_{2}$ & $32$ & $-7.964$ & $-7.230$ & $0.734$ \\
exact Faddeev~\cite{Nogga2003} & & & $-8.005$ & $-7.274$ & $0.731$ \\
experiment & & & $-8.482$ & $-7.718$ & $0.764$ \\
\bottomrule
\end{tabular}
\end{center}
\end{table}

Table~\ref{tab:a3} gives the splitting by model space and mesh. In the full model space of pair angular momenta $j\le4$, $40$ three-body channels, with the Fock block on the $pp$ $^{1}S_{0}$, $^{3}P_{0}$, $^{3}P_{1}$ and $^{3}P_{2}$--$^{3}F_{2}$ waves, the splitting is $0.734$~MeV at $N_{q}=32$; the $j\le4$ energies lie above the $S$-wave-pair ones because the separable $P$-wave representation reduces the binding, the truncation error documented in Ref.~\cite{NishonovValidation}. The Coulomb force in the $pp$ $P$ waves contributes $40$~keV to the splitting, the same at both meshes and converged in the Fock rank (Table~\ref{tab:rank}); eight Fock modes on the $^{1}D_{2}$ wave add a further $5$~keV, so the higher $pp$ waves contribute a few keV at most. Against the exact Faddeev energies of the CD~Bonn potential~\cite{Nogga2003} the separable representation gives a splitting $3$~keV above the exact one while it underbinds $^{3}$H by $41$~keV; with the omitted $pp$ $D$-wave Coulomb terms the deviation would be about $8$~keV, and deviations of this size lie within the residual of the separable representation. A rank-one three-nucleon force fitted to the experimental $^{3}$H energy changes $\Delta E$ by less than $10$~keV when its regulator is changed by a factor of two (Ref.~\cite{NishonovValidation}). The remainder against experiment, $30$~keV, is nearly the same ($33$~keV) for the exact calculation and lies in the charge-symmetry-breaking part beyond the charge dependence of the CD~Bonn potential, in agreement with the decomposition of the splitting into Coulomb and charge-asymmetric parts~\cite{Brandenburg1988}.

\subsection{The three-cluster nuclei}
\label{sec:a3a6}

The states are the ground states $^{3}$He($1/2^{+}$), $^{6}$He($0^{+}$), $^{6}$Li($1^{+}$), $^{9}$Be($3/2^{-}$), $^{9}$B($3/2^{-}$) and $^{12}$C($0^{+}$). The $\alpha\alpha$ interaction acts in the $L=0$, $2$ and $4$ waves, the $\alpha N$ interaction in the five waves $^{2}S_{1/2}$, $^{2}P_{1/2}$, $^{2}P_{3/2}$, $^{2}D_{3/2}$ and $^{2}D_{5/2}$, and the nucleon--nucleon input is that of Sec.~\ref{sec:a3}. Table~\ref{tab:3b} collects the three-body results with the Fock representation, from $A=3$, where the nuclear input is the most complete, to $A=6$, $A=9$ and $A=12$, where the cluster model is less complete and the Coulomb effects larger, each entry at the mesh and precision of Table~\ref{tab:precision}. The Coulomb displacement $\Delta E_{C}$ is defined per system. For $A=3$ it is $E(^{3}\mathrm{H})-E(^{3}\mathrm{He})$. For $^{6}$Li, $^{9}$Be and $^{12}$C it is the Coulomb-on minus the Coulomb-off energy, the Coulomb-off calculation projecting the forbidden states of the nuclear problem, so that each calculation is made with its own eigenstates; a definition with fixed vectors would differ by the last column of Table~\ref{tab:dressedtest}. For $^{6}$Li both calculations are in the particle basis, the Coulomb-off one with the Fock block removed. For the $A=9$ mirror pair it is $E(^{9}\mathrm{B})-E(^{9}\mathrm{Be})$. The experimental energies~\cite{Wang2021} are relative to the three-cluster threshold; the $^{6}$Li displacement has no experimental counterpart.

\begin{table}[htbp]
\caption{Three-body Coulomb benchmarks of the Fock representation (MeV) at the mesh of Table~\ref{tab:precision}; $N_{q}$ is the number of spectator-momentum mesh points, ``grid'' the piecewise grid of Appendix~\ref{sec:mesh} with $n_{g}=10$, and $\Delta E_{C}$ the Coulomb displacement defined in the text. The $^{6}$Li, $^{9}$Be and $^{9}$B rows carry all five $\alpha N$ waves ($^{2}S_{1/2}$, $^{2}P_{1/2}$, $^{2}P_{3/2}$, $^{2}D_{3/2}$, $^{2}D_{5/2}$) and the $^{6}$He row the $S$ and $P$ waves; the number in brackets in the $^{6}$He entry gives its change over the mesh range; \textsuperscript{a}~the $^{9}$B resonance position minus the $^{9}$Be energy, a resonance against a bound state (text).}
\label{tab:3b}
\begin{center}
\begin{tabular}{lccc}
\toprule
 & computed & mesh & experiment \\
\midrule
$E(^{3}\mathrm{H})$                    & $-7.964$ & $N_{q}=32$ & $-8.482$ \\
$E(^{3}\mathrm{He})$                   & $-7.230$ & $32$ & $-7.718$ \\
$\Delta E_{C}$ ($A=3$)                 & $0.734$  & $32$ & $0.764$  \\
\midrule
$E(^{6}\mathrm{He})$                   & $-0.908(1)$ & $20$--$32$ & $-0.975$ \\
$E(^{6}\mathrm{Li})$, Coulomb off      & $-4.243$ & $32$ & --       \\
$E(^{6}\mathrm{Li})$                   & $-3.262$ & $32$ & $-3.699$ \\
$\Delta E_{C}$ ($^{6}\mathrm{Li}$)     & $0.981$  & $32$ & --       \\
\midrule
$E(^{9}\mathrm{Be})$, Coulomb off      & $-4.871$ & $32$ & --       \\
$E(^{9}\mathrm{Be})$                   & $-2.041$ & $32$ & $-1.574$ \\
$\Delta E_{C}$ ($^{9}\mathrm{Be}$)     & $2.83$   & $32$ & --       \\
$E(^{9}\mathrm{B})$                    & $-0.292$ & $32$ & --       \\
$\Delta E_{C}$ ($A=9$ mirror)          & $1.749$  & $32$ & $1.851$\textsuperscript{a} \\
\midrule
$E(^{12}\mathrm{C})$, Coulomb off      & $-29.38$ & grid & --       \\
$E(^{12}\mathrm{C})$                   & $-19.96$ & grid & $-7.275$ \\
$\Delta E_{C}$ ($^{12}\mathrm{C}$)     & $9.42$   & grid & --       \\
\bottomrule
\end{tabular}
\end{center}
\end{table}

At $A=3$ the displacement is the $0.734$~MeV of Sec.~\ref{sec:a3}. At $A=6$, $^{6}$He has no Coulomb force and converges $67$~keV above experiment with two-body forces only. In $^{6}$Li the Fock block on the $\alpha p$ channels, with the dressed forbidden state, supplies the Coulomb force, absent in the model-input $^{6}$Li of Ref.~\cite{Nishonov2026a}. The $^{6}$Li Coulomb displacement of $0.98$~MeV, converged to a keV as a difference of two energies on the same mesh, lies slightly above the $0.84$--$0.95$~MeV Coulomb energies of the three-cluster models of Ref.~\cite{Schellingerhout1993}, where the Coulomb energy is shown to be perturbative. One systematic uncertainty of this comparison is not quantified at the three-body level: the $\alpha p$ Coulomb force is the point-charge one, whereas the folded force of those models is softer at short distances, and Ref.~\cite{FockI} notes that the folding, weaker than the point force at every radius, would lower the dressed $\alpha p$ $0s$ state; its size is not computed there. The remaining deficits against experiment are those of three-cluster models with two-body forces only.

The $\alpha\alpha$ systems have the largest Coulomb effects. The Coulomb displacement of $^{9}$Be is $2.8$~MeV, that of $^{12}$C $9.4$~MeV on the piecewise grid, about three times the $^{9}$Be value. Both are displacements of overbound model states, $^{12}$C by more than $10$~MeV (Sec.~\ref{sec:exact}), and a Coulomb displacement grows with the compactness of the state; they are properties of the model and not predictions for the nuclei.

The $A=9$ mirror pair tests the $\alpha p$ Coulomb treatment. $^{9}$B is $^{9}$Be with the neutron replaced by a proton and the Fock block on the five $\alpha p$ waves, the $^{2}S_{1/2}$ wave through the $\alpha p$ separable potential of Ref.~\cite{NishonovValidation} (the $\alpha n$ construction at the $\alpha p$ reduced mass) and the $P$ and $D$ waves through the $\alpha n$ separable potentials with the Fock terms added. Against $\alpha p$ potentials constructed in these waves too, the difference is a few keV in $^{6}$Li. The Fock block on $\alpha p$, the $\alpha\alpha$ Fock block common to both nuclei and the dressed $\alpha p$ forbidden state all enter the displacement. The mirror displacement is $1.75$~MeV, stable to a few keV from $N_{q}=22$ to $40$ ($1.749$~MeV at both $N_{q}=32$ and $40$). The same mirror pair has been studied in the $\alpha\alpha N$ model with the complex-scaling method, where the mirror-symmetry breaking is likewise attributed to the Coulomb interaction~\cite{Odsuren2026}, and a molecular-model estimate of the displacement from fitted $\alpha\alpha N$ densities gives $1.66$~MeV, the difference of the ground-state Coulomb energies of Ref.~\cite{Brooks2026}. In experiment $^{9}$B is unbound, a resonance $0.28$~MeV above the $\alpha\alpha p$ threshold, while the model binds it weakly, so the mirror displacement compares a bound-state energy with an experimental resonance position. The binding of $^{9}$B in the model reflects the overbinding of about $0.5$~MeV of the cluster input, which also makes $^{9}$Be too deep at the converged mesh; if it is common to the mirror pair it cancels in the displacement. The overbinding is known from $\alpha\alpha N$ models with pair forces fitted to the $\alpha N$ and $\alpha\alpha$ scattering data: three-body calculations of $^{9}$Be correct it with a repulsive phenomenological three-body force fitted to the ground state, of about $1.1$~MeV strength in Refs.~\cite{Casal2014,Kikuchi2016}; no such term is added here.

Every projection in Table~\ref{tab:3b} is the strict projection of Sec.~\ref{sec:kernel}.  Table~\ref{tab:lambda}, whose sequence is also the realistic-input illustration of Ref.~\cite{Nishonov2026a}, shows how the finite-strength pseudopotential approaches it in the four $A=6$ cases, with $\lambda=10^{2}$--$10^{5}$~MeV, on the same separable potentials as Table~\ref{tab:3b}. For the three cases without the Fock block the deviation from the strict projection decreases by up to an order of magnitude per decade of $\lambda$, and the two-point $1/\lambda$ extrapolation through the two largest values of $\lambda$, the slope $\kappa$ and the intercept of the last two columns, reproduces the strict value to well below $1$~keV. With the Fock block on, the two forbidden states of the system, the Coulomb-dressed $\alpha p$ and the nuclear $\alpha n$ $0s$, contribute to the $1/\lambda$ tail with opposite signs, so the sequence crosses zero and approaches the limit from above, with the positive slope of the table.  At $\lambda=10^{4}$~MeV the forbidden-state combination shifted by the pseudopotential to finite energy lies near the ground state (marked entry), and the extrapolation through the $3\times10^{4}$ and $10^{5}$ points again reproduces the strict value closely. The strict projection has no pseudostate: the Schur complement removes the forbidden block instead of shifting it.

\begin{table}[htbp]
\caption{Ground-state energy (MeV) at pseudopotential strength $\lambda$ (MeV) and in the strict projection. $\kappa$ and the intercept are the slope and the $\lambda\to\infty$ value of the fit $E(\lambda)=E(\infty)+\kappa/\lambda$ through the two largest values of $\lambda$ of each row; the asterisk marks the entry at which the pseudopotential-shifted forbidden state lies near the ground state (text). The $^{6}$Li isospin-basis row carries the $\alpha N$ $S$ and $P$ waves.}
\label{tab:lambda}
\begin{center}
\resizebox{\textwidth}{!}{%
\begin{tabular}{lccccccccc}
\toprule
system & $N_{q}$ & $10^{2}$ & $10^{3}$ & $10^{4}$ & $3\times10^{4}$ & $10^{5}$ & strict & $\kappa$ (MeV$^{2}$)\textsuperscript{a} & intercept \\
\midrule
$^{6}$He                        & 22 & $-0.93739$ & $-0.91609$ & $-0.90881$ & -- & $-0.90749$ & $-0.90733$ & $-14.7$ & $-0.90734$ \\
$^{6}$Li, isospin basis         & 26 & $-4.23713$ & $-4.15378$ & $-4.13632$ & -- & $-4.13381$ & $-4.13351$ & $-27.8$ & $-4.13354$ \\
$^{6}$Li, particle basis, Coulomb off & 26 & $-4.34926$ & $-4.26384$ & $-4.24640$ & -- & $-4.24388$ & $-4.24357$ & $-28.0$ & $-4.24360$ \\
$^{6}$Li, particle basis, Fock on & 26 & $-3.37921$ & $-3.27199$ & $-3.24175^{\ast}$ & $-3.2552$ & $-3.26001$ & $-3.26196$ & $+205$ & $-3.26206$ \\
\bottomrule
\end{tabular}}
\end{center}
\vspace{1pt}
{\footnotesize\raggedright\noindent\textsuperscript{a}\,$\kappa<0$ means that the sequence reaches the limit from below, $\kappa>0$ from above.\par}
\end{table}

\section{Deep versus oscillator forbidden states}
\label{sec:prescription}

Two prescriptions for the projected function are in use, so the eigenstate condition can be tested directly. The \emph{deep-eigenstate} prescription takes the forbidden state as the bound state of the separable potential itself, Coulomb-dressed where the Coulomb force is on; it satisfies the condition by construction. The \emph{harmonic-oscillator} prescription of the orthogonality-condition model takes the microscopic oscillator functions, with $\nu_{\rm rel}=0.5625$~fm$^{-2}$ for the $\alpha\alpha$ relative motion and the oscillator length $b=1.36$~fm for $\alpha N$ (Ref.~\cite{FockI}, Appendix~B). These functions are close to the eigenstates of the interactions used here but not identical to them, so the elimination removes a slightly different state.

That the two choices differ in the $3\alpha$ system is known. Fujiwara, Kohno and Suzuki~\cite{Fujiwara2004OCM} found $2$~MeV in the $^{12}$C ground state without the Coulomb force, the oscillator state being the less bound. Tursunov and Mazumdar~\cite{Tursunov2022} found a strong sensitivity of the $0^{+}$ and $2^{+}$ states to the elimination of the forbidden states, with spurious deep states produced by the near-critical eigenstates of the Pauli projector. Moriya, Horiuchi and Zhou~\cite{Moriya2023} build a coordinate-space basis which is nearly orthogonal to the oscillator forbidden states by construction, so that the pseudopotential is not needed (negligible residual forbidden component, basis reduced from about $1000$ to about $100$ functions); that approach modifies the basis, the present one the operator, with one Schur complement per pair. Deltuva~\cite{Deltuva2026} has compared the projection with the supersymmetric transformation~\cite{Baye1987} for core-plus-two-nucleon systems, where the forbidden states are the deep eigenstates of the core--nucleon potential as here and an oscillator approximation is reported to make little difference; the $\alpha\alpha$ pair, where the difference is largest here, does not occur there.

\subsection{The three-cluster binding energies}
\label{sec:conv3b}

In the test the deep eigenstate is replaced by the oscillator function, with everything else kept fixed, in $A=6$, which has no $\alpha\alpha$ pair, in $^{9}$Be, where the replaced state is Coulomb-dressed, and in $^{12}$C; the two-body form of the test, on the phase shifts, is in Ref.~\cite{FockI}. Table~\ref{tab:prescription} gives the three-body comparison, in which only the forbidden state changes from row to row. The $A=6$ rows are nuclear-only calculations in the isospin basis with the $\alpha N$ $S$ and $P$ waves; the $^{9}$Be and $^{12}$C rows have the $\alpha\alpha$ Fock block on, with the BFW $\alpha\alpha$ separable potential used throughout this paper. The shifts converge in the mesh before the energies do (Table~\ref{tab:meshshift} of Appendix~\ref{sec:appB}).

\begin{table}[htbp]
\caption{Ground-state energies (MeV) with the deep eigenstates (deep) and the oscillator functions (h.o.) as forbidden states, from otherwise identical calculations; the shift is h.o.\ minus deep. The $^{6}$He and $^{9}$Be rows are at $N_{q}=32$, the $^{6}$Li rows at $N_{q}=26$, the $^{12}$C Coulomb-off row on the piecewise grid with $n_{g}=10$ and the Coulomb-on row with $n_{g}=20$; $n_{\alpha\alpha}$ is the number of $\alpha\alpha$ pairs. For the $\alpha\alpha$ systems both calculations are given with the $\alpha\alpha$ Coulomb force off and on. The $^{6}$Li isospin rows carry the $\alpha N$ $S$ and $P$ waves.}
\label{tab:prescription}
\begin{center}
\begin{tabular}{llcrrr}
\toprule
system & Coulomb & $n_{\alpha\alpha}$ & deep & h.o.\ & shift \\
\midrule
$^{6}$He              & off & 0 & $-0.9076$ & $-0.8721$ & $+36$~keV \\
$^{6}$Li (isospin)    & off & 0 & $-4.1335$ & $-4.2442$ & $-111$~keV \\
$^{9}$Be              & off & 1 & $-4.871$  & $-5.031$  & $-0.16$~MeV \\
                      & on  & 1 & $-2.041$  & $-2.464$  & $-0.42$~MeV \\
$^{12}$C, $0^{+}$     & off & 3 & $-29.38$  & $-28.88$  & $+0.50$~MeV \\
                      & on  & 3 & $-19.91$  & $-17.62$  & $+2.29$~MeV \\
\bottomrule
\end{tabular}
\end{center}
\end{table}

The $A=6$ shifts have opposite signs, and the $^{6}$Li shift exceeds the $52$~keV spread of the $^{6}$Li energy between the two $\alpha N$ fits of Ref.~\cite{NishonovValidation}; the prescription shift therefore depends on the system and does not cancel between systems. In the $3\alpha$ system the oscillator choice is the less bound at every mesh, as in the $3\alpha$ calculations of Fujiwara, Kohno and Suzuki~\cite{Fujiwara2004OCM} without the Coulomb force and of Matsumura \emph{et al.}~\cite{Matsumura2006} with it. The sign agrees with the literature while the size differs: with the Coulomb force on the shift is $2.29$~MeV here against $0.65$~MeV in Ref.~\cite{Matsumura2006}, and with the Coulomb force off $0.50$~MeV against $2$~MeV in Ref.~\cite{Fujiwara2004OCM}. In $^{9}$Be, for which no oscillator-prescription value appears to have been published, the oscillator choice is the more bound, and the shifts are not monotone in $n_{\alpha\alpha}$. With the $\alpha\alpha$ Coulomb force off the substitution changes both nuclei by a fraction of an MeV; with it on, $^{12}$C changes by $2.29$~MeV. The oscillator function approximates the nuclear eigenstate closely (overlaps with the Coulomb-off eigenstates between $0.98$ and $1$, Table~\ref{tab:pairspec}), but it is not Coulomb-dressed, and the Coulomb dressing of the projected states alone accounts for most of the difference between the two $^{12}$C shifts (Sec.~\ref{sec:exact}, Table~\ref{tab:dressedtest}).

The mechanism is analysed in Appendix~\ref{sec:appB}. The $^{9}$Be shift comes mainly from the $1s$ forbidden state, whose node structure no single oscillator function reproduces (Table~\ref{tab:rows}), and the projected $\alpha\alpha$ pair keeps only its shallow $^{8}$Be state at $\nu_{\rm rel}$ but a deep spurious state at the single-$\alpha$ width (Table~\ref{tab:pairspec}), so the oscillator state should be taken at $\nu_{\rm rel}$ (Appendix~\ref{sec:appB}). In the $^{6}$Li particle basis the oscillator function in place of the dressed $\alpha p$ state gives $-74$~keV and in place of the nuclear $\alpha n$ state $-92$~keV more (Table~\ref{tab:li6}), so the part specific to the Coulomb dressing is about $20$~keV (Appendix~\ref{sec:appB}). The $\alpha\alpha$ pair, with phase-shift differences of several degrees on shell (Ref.~\cite{FockI}) and $0.4$~MeV in $^{9}$Be, is where the distinction between the dressed and the oscillator state matters.

This does not rank the deep eigenstate above the oscillator function in general. The forbidden states are those of the interaction used: for a phenomenological cluster potential, its own deep bound states dressed by the long-range interaction included in the calculation; for a resonating-group kernel, the microscopic forbidden states adapted to that operator. This agrees with the recommendation of Ref.~\cite{Matsumura2006}, made in a different formalism.

\section{Summary and outlook}
\label{sec:summary}

A bound-state Faddeev kernel samples only negative pair energies, where the Fock representation of the Coulomb interaction is exact. The Coulomb force of every charged pair therefore enters the momentum-space kernel as a convergent separable block beside the nuclear terms, without screening or distortion of the form factors; only the rank of the pair interaction increases. The Feshbach--Schur projection, in the Schur-complement form of Ref.~\cite{Nishonov2026a}, acts on the Pauli-forbidden states of the charged pairs as eigenstates of the pair Hamiltonian which that kernel contains, the Coulomb-dressed states, several per pair at once; the present results rest on this combination.

With the Fock block the $^{3}$H--$^{3}$He splitting of the CD~Bonn potential agrees with the exact Faddeev result to $3$~keV, the $pp$ $P$ waves contributing $40$~keV of it, while in the $S$-wave model space the Coulomb-distorted form factors give $0.37$ against $0.69$~MeV (Table~\ref{tab:he3}). The $^{6}$Li Coulomb displacement lies close to those of other three-cluster models, and with the block on the $\alpha p$ and $\alpha\alpha$ pairs the Coulomb displacements of the three-cluster nuclei follow from one separable input: $2.8$~MeV in $^{9}$Be, $1.75$~MeV for the $^{9}$Be--$^{9}$B mirror pair and $9.4$~MeV in $^{12}$C on the piecewise grid. The Fock rank of $20$ is converged to better than a keV in $^{3}$He and, at the finest mesh, in $^{9}$Be, and to $3$~keV in $^{12}$C at $n_{g}=20$.

The choice of the projected state shifts the three-body energies as follows.  Projecting the undressed eigenstate instead of the Coulomb-dressed one changes $^{9}$Be by $60$~keV and $^{12}$C by $1.63$~MeV, where the projected states enter through three pairs and the binding is ten times larger. The oscillator function of the usual prescription shifts $^{6}$He by $36$~keV and $^{6}$Li by a tenth of an MeV, with opposite signs. With the Coulomb force on it shifts $^{9}$Be by $0.42$~MeV and $^{12}$C by $2.29$~MeV; with the Coulomb force off the same replacement changes them by $0.16$ and $0.50$~MeV, so in $^{12}$C the two prescriptions differ mainly through the Coulomb dressing of the projected states.

Where deep forbidden states are removed in the presence of a Coulomb force, the eigenstate condition requires the projected states to be the Coulomb-dressed ones. The same distinction may apply to other cluster or halo calculations which remove deep forbidden states of charged pairs, by projection or by a supersymmetric transformation~\cite{Baye1987} such as that of Ref.~\cite{Deltuva2026}; its size has not been evaluated there, and in the systems of this paper it ranges from about $20$~keV, as estimated for the $\alpha N$ pairs in Sec.~\ref{sec:conv3b}, to the $3\alpha$ value of Table~\ref{tab:dressedtest}.

The same kernel, with the pair $\tau$~matrices at the same negative pair energies, describes elastic scattering below the breakup threshold. Elastic scattering of a charged particle on a charged bound pair, $p+d$ in the three-nucleon system and $\alpha+d$ on the $^{6}$Li kernel of this paper, is the application that follows.

\appendix

\section{Numerical details: spectator mesh and Fock rank}
\label{sec:mesh}

The spectator momentum is discretized on a Gauss--Legendre mesh of $N_{q}$ points mapped by $q=q_{0}\,(1/x-1)$ with $q_{0}=0.6$~fm$^{-1}$, called the mapped mesh; half of its points lie below $0.6$~fm$^{-1}$. For $^{12}$C a piecewise Gauss--Legendre grid of the kind used by Fujiwara, Kohno and Suzuki~\cite{Fujiwara2004OCM} is used as well, called the piecewise grid: $n_{g}$ points on each of the intervals $0$--$1$, $1$--$3$ and $3$--$6$~fm$^{-1}$ and five mapped points beyond.

The momentum scale of a state bound by $E$ in the $\alpha$ plus pair channel is $\sqrt{2\bar{\mu}_a|E|}$, with $\bar{\mu}_a$ the spectator reduced mass of Sec.~\ref{sec:faddeev}: about $0.5$~fm$^{-1}$ for $^{9}$Be and $1.6$~fm$^{-1}$ for the $3\alpha$ state, which is bound by $20$~MeV in this model. The $^{9}$Be amplitude is thus concentrated where the mapped mesh is dense and the $^{12}$C amplitude where it is sparse.

\begin{table}[!htb]
\caption{The $^{12}$C ground-state energy (MeV), with the Coulomb force on and the Coulomb-dressed forbidden states projected, against the spectator mesh: the mapped mesh with $q_{0}=0.6$~fm$^{-1}$ and $N_{q}$ points, and the piecewise grid with $n_{g}$ points per interval, both defined in the text; the third column is the fraction of mesh points above $1$~fm$^{-1}$.}
\label{tab:mesh12c}
\begin{center}
\begin{tabular}{llcc}
\toprule
mesh & & points above $1$~fm$^{-1}$ & $E(^{12}\mathrm{C})$ \\
\midrule
mapped, $q_{0}=0.6$    & $N_{q}=32$ & $38\%$ & $-20.421$ \\
                   & $N_{q}=40$ & $38\%$ & $-20.050$ \\
                   & $N_{q}=48$ & $38\%$ & $-19.970$ \\
\midrule
piecewise          & $n_{g}=10$ ($35$ points) & $71\%$ & $-19.964$ \\
                   & $n_{g}=15$ ($50$ points) & $70\%$ & $-19.917$ \\
                   & $n_{g}=20$ ($65$ points) & $69\%$ & $-19.910$ \\
\bottomrule
\end{tabular}
\end{center}
\end{table}

Table~\ref{tab:mesh12c} shows the consequence for $^{12}$C: on the mapped mesh the energy converges slowly with $N_{q}$, while the $n_{g}=10$ grid energy already equals that of the finest mapped mesh to a few keV, and the grid leaves $^{9}$Be at its mapped-mesh value (Table~\ref{tab:3b}). The steps between the grid rows of Table~\ref{tab:mesh12c} place the $n_{g}=10$ entries of this paper about $55$~keV below the converged value (Table~\ref{tab:precision}).

The differences of Tables~\ref{tab:dressedtest} and~\ref{tab:meshshift} converge before the energies do, the projected forbidden states being confined to momenta which every mesh resolves.

Table~\ref{tab:rank} gives the variation of the Fock rank in the three-body calculations themselves. For the $pp$ pairs in $^{3}$He the energy is converged in $N_{C}$ to better than a keV. For the $\alpha\alpha$ pairs the rank and the spectator mesh are coupled. The Fock modes of high order oscillate in the pair momentum and require more spectator points than the nuclear form factors do, so the step from $N_{C}=20$ to $30$ in $^{9}$Be becomes negligible only at the finest mesh, where the $N_{C}=20$ energy itself no longer moves.  The expansion is therefore converged at $N_{C}=20$ once the mesh resolves the modes, and the residuals of the tabulated $^{9}$Be and $^{12}$C entries which follow from Table~\ref{tab:rank} are those of Table~\ref{tab:precision}.

\begin{table}[htbp]
\caption{Convergence of the three-body ground-state energies in the Fock rank $N_{C}$. Each entry is the change of the energy (keV) under the stated variation  at the stated mesh (meshes as in Appendix~\ref{sec:mesh}). The $pp$ $P$-wave entry is the spread of the energies at $N_{C}=10$, $20$ and $22$.  The last two $^{9}$Be rows vary the mesh at fixed rank.}
\label{tab:rank}
\begin{center}
\begin{tabular}{llllc}
\toprule
system & Fock block on & mesh & variation & change (keV) \\
\midrule
$^{3}$He  & $pp$ $^{1}S_{0}$ & $N_{q}=16$ & $N_{C}$: $10\to20$ & $0.8$ \\
          &                  & $16$       & $N_{C}$: $20\to22$ & $0.07$ \\
          & $pp$ $^{3}P_{0}$, $^{3}P_{1}$, $^{3}P_{2}$--$^{3}F_{2}$ & $16$ & $N_{C}=10$, $20$, $22$ & $0.2$ \\
\midrule
$^{9}$Be  & $\alpha\alpha$   & $N_{q}=32$ & $N_{C}$: $20\to30$ & $12$ \\
          &                  & $40$       & $N_{C}$: $20\to30$ & $3$ \\
          &                  & $48$       & $N_{C}$: $20\to30$ & $0.6$ \\
          &                  & $N_{C}=20$ & $N_{q}$: $32\to40$ & $2.8$ \\
          &                  & $N_{C}=20$ & $N_{q}$: $40\to48$ & $0$ \\
\midrule
$^{12}$C  & $\alpha\alpha$   & grid, $n_{g}=10$ & $N_{C}$: $20\to30$ & $18$ \\
          &                  & grid, $n_{g}=20$ & $N_{C}$: $20\to30$ & $3$ \\
\bottomrule
\end{tabular}
\end{center}
\end{table}

\section{Details of the oscillator comparison}
\label{sec:appB}

Table~\ref{tab:meshshift} gives the convergence of the prescription shifts of Table~\ref{tab:prescription} in the mesh: the $A=6$ and $^{9}$Be shifts are stable from $N_{q}=22$ on while the energies themselves still move, and the $^{12}$C shift on the piecewise grid changes by $7$~keV in its last step.

\begin{table}[htbp]
\caption{Mesh dependence of the prescription shift (h.o.\ minus deep) of Table~\ref{tab:prescription}; keV for $A=6$, MeV for $^{9}$Be and $^{12}$C; ``grid'' is the piecewise grid of Appendix~\ref{sec:mesh} with $n_{g}$ points per interval.}
\label{tab:meshshift}
\begin{center}
\footnotesize
\begin{tabular}{lcccccccc}
\toprule
system & $N_{q}=16$ & $22$ & $26$ & $32$ & $40$ & grid, $n_{g}=10$ & $15$ & $20$ \\
\midrule
$^{6}$He (keV)              & $+35.0$ & $+35.8$ & $+35.5$ & $+35.5$ & -- & -- & -- & -- \\
$^{6}$Li isospin (keV)      & $-105.6$ & $-109.1$ & $-110.7$ & -- & -- & -- & -- & -- \\
$^{9}$Be (MeV)              & $-0.24$ & $-0.42$ & $-0.42$ & $-0.42$ & -- & -- & -- & -- \\
$^{12}$C (MeV)              & $+2.2$ & $+3.8$ & $+3.5$ & $+2.78$ & $+2.44$ & $+2.346$ & $+2.293$ & $+2.286$ \\
\bottomrule
\end{tabular}
\end{center}
\end{table}

The $^{9}$Be shift can be traced to one state of the oscillator set by exchanging the states one at a time (Table~\ref{tab:rows}). The deep $1s$ eigenstate in place of the oscillator $1s$, with the oscillator $0s$ and $0d$ kept, returns $^{9}$Be to the deep value, while the oscillator $1s$ alone gives most of the shift, at $\nu_{\rm rel}$ and at its width of maximal overlap with the deep state (last column of Table~\ref{tab:rows}); the $0d$ function at its width of maximal overlap moves $^{9}$Be slightly away from the deep value. The projected $\alpha\alpha$ $L=0$ pair keeps only its shallow state in every case. No single oscillator function reproduces its node structure at either width of Table~\ref{tab:rows}, and its overlap deficit of under two percent accounts for most of the $^{9}$Be shift. The $0s$ and $0d$ functions produce no comparable shift.

\begin{table}[htbp]
\caption{$^{9}$Be with the $\alpha\alpha$ forbidden states exchanged one at a time (MeV), all rows at $N_{q}=32$ with the $\alpha\alpha$ Coulomb force on, the deep states being the Coulomb-dressed eigenstates; $\nu$ is the oscillator width (fm$^{-2}$), and the last column is the shift of each row against the first as a fraction of the full oscillator shift.}
\label{tab:rows}
\begin{center}
\small
\setlength{\tabcolsep}{4pt}
\begin{tabular}{lcc}
\toprule
forbidden states & $E(^{9}\mathrm{Be})$ & fraction of the full shift \\
\midrule
all deep eigenstates                                  & $-2.041$ & $0$ \\
all oscillator, $\nu_{\rm rel}$                       & $-2.464$ & $1$ \\
oscillator, $0d$ at its width of maximal overlap, $\nu=0.45$    & $-2.477$ & $1.03$ \\
deep $1s$; oscillator $0s$, $0d$                     & $-2.046$ & $0.01$ \\
deep $0s$, $0d$; oscillator $1s$                     & $-2.339$ & $0.70$ \\
oscillator $1s$ at its width of maximal overlap, $\nu=0.54$     & $-2.317$ & $0.65$ \\
\bottomrule
\end{tabular}
\end{center}
\end{table}

The bound-state spectrum of the projected pair interaction, Eq.~(\ref{eq:taueff}), with the respective forbidden states removed, shows directly what each prescription removes. Table~\ref{tab:pairspec} gives it together with the overlaps of the oscillator functions with the deep forbidden states, at the width of the relative motion $\nu_{\rm rel}$ and at the single-$\alpha$ width $0.28125$~fm$^{-2}$; the $\alpha N$ overlap is $0.989$ at $b=1.36$~fm. With the Coulomb force the shallow $L=0$ state becomes the $^{8}$Be resonance when the dressed eigenstates are projected, while the projection of the oscillator functions leaves a bound state at $-0.31$~MeV (Ref.~\cite{FockI}); without it, the oscillator model of Ref.~\cite{Fujiwara2004OCM} binds the $2\alpha$ pair at $-1.91$~MeV (their Table~4, Coulomb force off).  The deep pole is removed with either vector in every channel, as the theorem of Ref.~\cite{Fujiwara2004OCM} states for the oscillator vector; what remains depends on the vector.  With the deep eigenstates, or with the oscillator functions at $\nu_{\rm rel}$, the $\alpha N$ pair has no bound state and the $\alpha\alpha$ $L=0$ pair keeps only its shallow $^{8}$Be state. At the single-$\alpha$ width the overlaps decrease and the projected $\alpha\alpha$ pair keeps a deep spurious state. The model forbidden state should therefore be taken at the width of the relative coordinate of the channel it is projected from, and the spectrum of the projected pair should be checked before a three-body calculation is built on it.

\begin{table}[htbp]
\caption{Bound states of the $\alpha\alpha$ pair (MeV, Coulomb force off; the deep forbidden states are here the eigenstates of the Coulomb-off pair), unprojected and with the strict projection of the deep forbidden states or of the oscillator functions at the two widths, and the overlaps of the oscillator functions with the deep states.}
\label{tab:pairspec}
\begin{center}
\begin{tabular}{lcc}
\toprule
 & $L=0$ & $L=2$ \\
\midrule
unprojected                            & $-80.29$, $-31.13$, $-2.25$ & $-25.87$ \\
deep forbidden states projected       & $-2.25$ & none \\
oscillator, $\nu_{\rm rel}$, projected & $-2.51$ & none \\
oscillator, single-$\alpha$ width, projected & $-16.3$ & $-4.3$ \\
overlap, $\nu_{\rm rel}$ ($0s$, $1s$; $0d$) & $0.996$, $0.983$ & $0.990$ \\
overlap, single-$\alpha$ width            & $0.886$, $0.720$ & $0.883$ \\
\bottomrule
\end{tabular}
\end{center}
\end{table}

\begin{table}[!htb]
\caption{$^{6}$Li in the particle basis with the Fock block on all five $\alpha p$ waves and the strict projection, at the converged mesh (MeV): the ground-state energy with the deep Coulomb-dressed forbidden states, and with the oscillator function in place of the deep state in the channels listed; the shift is the difference against the first row, and the last column the shift added by each channel.}
\label{tab:li6}
\begin{center}
\begin{tabular}{lccc}
\toprule
forbidden states & $E(^{6}\mathrm{Li})$ & shift (keV) & per channel (keV) \\
\midrule
deep, dressed, in all channels                          & $-3.2620$ & -- & -- \\
oscillator in the dressed $\alpha p$ $^{2}S_{1/2}$ channel  & $-3.3357$ & $-74$ & $-74$ \\
oscillator in the $\alpha p$ and the $\alpha n$ channel & $-3.4282$ & $-166$ & $-92$ \\
\bottomrule
\end{tabular}
\end{center}
\end{table}

In $^{6}$Li the $\alpha N$ forbidden state is Coulomb-dressed in the particle basis, and there the condition can be tested with the same Hamiltonian and only the projected vector changed. With the Coulomb force off this basis binds $^{6}$Li more than the isospin basis (Table~\ref{tab:lambda}), the two differing in the $\alpha N$ model space, the $S$ and $P$ waves in the isospin basis against all five waves in the particle basis. Table~\ref{tab:li6} gives the test: three separable potentials, identical except for the forbidden states, with the Fock block on all five $\alpha p$ waves and the strict projection everywhere, at $N_{q}=26$ (Table~\ref{tab:precision}). The oscillator state is the more bound in both replacements, as in the isospin basis (Table~\ref{tab:prescription}). The two channels contribute shifts of the same order (last column of Table~\ref{tab:li6}), so the part specific to the Coulomb dressing is of the order of their difference, about $20$~keV, if the two channels contribute alike apart from the dressing.

\bibliographystyle{unsrt}
{\footnotesize
\bibliography{refs}

@article{Fock1935,
  author  = {V. Fock},
  title   = {{Zur Theorie des Wasserstoffatoms}},
  journal = {Z. Phys.},
  volume  = {98},
  pages   = {145},
  year    = {1935}
}

@article{Alt1978,
  author  = {E. O. Alt and W. Sandhas and H. Ziegelmann},
  title   = {{C}oulomb effects in three-body reactions with two charged particles},
  journal = {Phys. Rev. C},
  volume  = {17},
  pages   = {1981},
  year    = {1978}
}

@article{vanHaeringen1982,
  author  = {L. P. Kok and H. van Haeringen},
  title   = {Bound-state solution in momentum space of three-particle equations with {C}oulomb interaction},
  journal = {Czech. J. Phys. B},
  volume  = {32},
  pages   = {311},
  year    = {1982}
}

@article{PappPlessas1996,
  author  = {Z. Papp and W. Plessas},
  title   = {{C}oulomb-{S}turmian separable expansion approach: {T}hree-body {F}addeev calculations for {C}oulomb-like interactions},
  journal = {Phys. Rev. C},
  volume  = {54},
  pages   = {50},
  year    = {1996}
}

@misc{Nishonov2026a,
  author  = {M. M. Nishonov},
  title   = {From {O}rthogonalizing {P}seudopotential to the {F}eshbach--{S}chur {P}rojection},
  note    = {arXiv:2603.03842},
  year    = {2026}
}

@article{Alt2002,
  author  = {E. O. Alt and A. M. Mukhamedzhanov and M. M. Nishonov and A. I. Sattarov},
  title   = {Proton-deuteron elastic scattering from 2.5 to 22.7 {MeV}},
  journal = {Phys. Rev. C},
  volume  = {65},
  pages   = {064613},
  year    = {2002}
}

@article{Machleidt2001,
  author  = {R. Machleidt},
  title   = {High-precision, charge-dependent {Bonn} nucleon-nucleon potential},
  journal = {Phys. Rev. C},
  volume  = {63},
  pages   = {024001},
  year    = {2001}
}

@article{Schellingerhout1993,
  author  = {N. W. Schellingerhout and L. P. Kok and S. A. Coon and R. M. Adam},
  title   = {Nucleon polarization in three-body models of polarized $^{6}$Li},
  journal = {Phys. Rev. C},
  volume  = {48},
  pages   = {2714},
  year    = {1993}
}

@article{EST1973,
  author  = {D. J. Ernst and C. M. Shakin and R. M. Thaler},
  title   = {Separable representations of two-body interactions},
  journal = {Phys. Rev. C},
  volume  = {8},
  pages   = {46},
  year    = {1973}
}

@article{Matsumura2006,
  author  = {H. Matsumura and M. Orabi and Y. Suzuki and Y. Fujiwara},
  title   = {Removal of forbidden states in a three-alpha system},
  journal = {Nucl. Phys. A},
  volume  = {776},
  pages   = {1},
  year    = {2006}
}

@article{Fujiwara2004OCM,
  author  = {Y. Fujiwara and M. Kohno and Y. Suzuki},
  title   = {Solving three-cluster {OCM} equations in the {F}addeev formalism},
  journal = {Few-Body Syst.},
  volume  = {34},
  pages   = {237},
  year    = {2004}
}

@article{Brandenburg1988,
  author  = {R. A. Brandenburg and G. S. Chulick and Y. E. Kim and D. J. Klepacki and R. Machleidt and A. Picklesimer and R. M. Thaler},
  title   = {Nuclear charge symmetry breaking and the $^{3}${H}--$^{3}${He} binding energy difference},
  journal = {Phys. Rev. C},
  volume  = {37},
  pages   = {781},
  year    = {1988}
}

@article{Tursunov2022,
  author  = {E. M. Tursunov and I. Mazumdar},
  title   = {Peculiarity of the $^{12}${C}$(0^+)$ and $^{12}${C}$(2^+)$ energy spectrum in a $3\alpha$ model},
  journal = {Phys. At. Nucl.},
  volume  = {85},
  pages   = {160},
  year    = {2022},
  note    = {arXiv:2001.08995}
}

@article{Moriya2023,
  author  = {H. Moriya and W. Horiuchi and B. Zhou},
  title   = {Novel approach to the removal of the {P}auli-forbidden states in the orthogonality condition model: a case of multi-$\alpha$ systems},
  journal = {Eur. Phys. J. A},
  volume  = {59},
  pages   = {197},
  year    = {2023},
  note    = {arXiv:2308.08749}
}

@article{Deltuva2005,
  author  = {A. Deltuva and A. C. Fonseca and P. U. Sauer},
  title   = {Momentum-space treatment of the {C}oulomb interaction in three-nucleon reactions with two protons},
  journal = {Phys. Rev. C},
  volume  = {71},
  pages   = {054005},
  year    = {2005}
}

@article{KKNN1979,
  author  = {Kanada, H. and Kaneko, T. and Nagata, S. and Nomoto, M.},
  title   = {{Microscopic study of nucleon-$^4$He scattering and effective nuclear potentials}},
  journal = {Prog. Theor. Phys.},
  volume  = {61},
  pages   = {1327},
  year    = {1979}
}

@article{BFW1977,
  author  = {Buck, B. and Friedrich, H. and Wheatley, C.},
  title   = {{Local potential models for the scattering of complex nuclei}},
  journal = {Nucl. Phys. A},
  volume  = {275},
  pages   = {246},
  year    = {1977}
}

@article{AGS1967,
  author  = {Alt, E. O. and Grassberger, P. and Sandhas, W.},
  title   = {{Reduction of the three-particle collision problem to multi-channel two-particle Lippmann-Schwinger equations}},
  journal = {Nucl. Phys. B},
  volume  = {2},
  pages   = {167},
  year    = {1967}
}

@article{Lovelace1964,
  author  = {Lovelace, C.},
  title   = {{Practical theory of three-particle states. I. Nonrelativistic}},
  journal = {Phys. Rev.},
  volume  = {135},
  pages   = {B1225},
  year    = {1964}
}

@article{Deltuva2026,
  author  = {Deltuva, A.},
  title   = {{Comparison of Pauli projection and supersymmetric transformation methods for three-body nuclear structure and reactions}},
  journal = {Phys. Rev. C},
  volume  = {113},
  pages   = {034302},
  year    = {2026},
  note    = {arXiv:2602.16674},
  doi     = {10.1103/k962-3236}
}

@article{Brooks2026,
  author  = {Brooks, A. D. and Bishop, J. and Kokalova, Tz. and Pirrie, S. and Wheldon, C.},
  title   = {{Coulomb energy differences in $^{9}$B using a molecular model}},
  journal = {J. Phys. G},
  volume  = {53},
  pages   = {025103},
  year    = {2026},
  doi     = {10.1088/1361-6471/ae42d8}
}

@article{Odsuren2026,
  author  = {Odsuren, M. and Kikuchi, Y. and Myo, T. and Kat{\=o}, K.},
  title   = {{Mirror symmetry and cluster structures of resonant states in $^{9}$Be and $^{9}$B}},
  journal = {Phys. Rev. C},
  volume  = {113},
  pages   = {024324},
  year    = {2026},
  doi     = {10.1103/2kn8-vd5t}
}

@article{Casal2014,
  author  = {Casal, J. and Rodr{\'i}guez-Gallardo, M. and Arias, J. M. and Thompson, I. J.},
  title   = {{Astrophysical reaction rate for $^{9}$Be formation within a three-body approach}},
  journal = {Phys. Rev. C},
  volume  = {90},
  pages   = {044304},
  year    = {2014}
}

@article{Kikuchi2016,
  author  = {Kikuchi, Y. and Odsuren, M. and Myo, T. and Kat{\=o}, K.},
  title   = {{Photodisintegration cross section of $^{9}$Be up to 16 MeV in the $\alpha+\alpha+n$ three-body model}},
  journal = {Phys. Rev. C},
  volume  = {93},
  pages   = {054605},
  year    = {2016}
}

@misc{NishonovValidation,
  author = {M. M. Nishonov},
  title  = {{Separable representations of two-body interactions for Faddeev calculations of light nuclei}},
  note   = {arXiv:2609.08775},
  year   = {2026}
}

@article{Kukulin1974,
  author = {Krasnopol'skii, V. M. and Kukulin, V. I.},
  title = {{Exclusion of occupied states from the nuclear optical amplitudes and potentials}},
  journal = {Yad. Fiz.},
  volume = {20},
  pages = {883},
  year = {1974},
  note = {English translation: Sov. J. Nucl. Phys. 20, 470 (1975)}
}

@article{Saito1969,
  author  = {Saito, S.},
  title   = {{Interaction between clusters and {P}auli principle}},
  journal = {Prog. Theor. Phys.},
  volume  = {41},
  pages   = {705},
  year    = {1969},
  doi     = {10.1143/PTP.41.705}
}

@misc{FockI,
  author = {M. M. Nishonov},
  title  = {{The {F}ock representation of the {C}oulomb interaction. {I}. {T}wo-body problem}},
  note   = {arXiv:2609.16585},
  year   = {2026}
}

@article{Baye1987,
  author  = {Baye, D.},
  title   = {{Supersymmetry between deep and shallow nucleus-nucleus potentials}},
  journal = {Phys. Rev. Lett.},
  volume  = {58},
  pages   = {2738},
  year    = {1987},
  doi     = {10.1103/PhysRevLett.58.2738}
}

@article{Nogga2003,
  author  = {Nogga, A. and Kievsky, A. and Kamada, H. and Gl{\"o}ckle, W. and Marcucci, L. E. and Rosati, S. and Viviani, M.},
  title   = {{The three-nucleon bound state using realistic potential models}},
  journal = {Phys. Rev. C},
  volume  = {67},
  pages   = {034004},
  year    = {2003}
}

@article{Wang2021,
  author  = {Wang, M. and Huang, W. J. and Kondev, F. G. and Audi, G. and Naimi, S.},
  title   = {{The {AME} 2020 atomic mass evaluation ({II}). {T}ables, graphs and references}},
  journal = {Chin. Phys. C},
  volume  = {45},
  pages   = {030003},
  year    = {2021}
}

@article{Kharchenko2018,
  author  = {V. F. Kharchenko},
  title   = {Partial-wave {C}oulomb transition matrices for attractive interaction by {F}ock's method},
  journal = {J. Math. Phys.},
  volume  = {59},
  pages   = {012202},
  year    = {2018},
  note    = {arXiv:1610.01431}
}

@misc{Kharchenko2017a,
  author  = {V. F. Kharchenko},
  title   = {Partial-wave {C}oulomb $t$-matrices for like-charged particles at ground-state energy},
  note    = {arXiv:1708.06109},
  year    = {2017}
}

@misc{Kharchenko2017b,
  author  = {V. F. Kharchenko},
  title   = {Analytical solution of the integral equation for partial wave {C}oulomb $t$-matrices at excited-state energy},
  note    = {arXiv:1709.00897},
  year    = {2017}
}

@misc{Kharchenko2019,
  author  = {V. F. Kharchenko},
  title   = {{C}oulomb transition matrix at negative energy and integer values of interaction parameter},
  note    = {arXiv:1905.13479},
  year    = {2019}
}
}
\end{document}